\documentclass[conference]{IEEEtran}
\IEEEoverridecommandlockouts

\usepackage{amsmath,graphicx}

\usepackage{amssymb}

\usepackage{tikz}
\usepackage{pgfplots}
\pgfplotsset{compat=1.3}% <-- moves axis labels near ticklabels (respects tick label widths)
\usetikzlibrary{arrows,snakes,shapes}

\usepackage{color}

\DeclareMathAlphabet{\mathbit}{OML}{cmr}{bx}{it}

\DeclareMathOperator{\Q}{Q}
\DeclareMathOperator{\E}{E}

\newcommand\Sign{\operatorname{sign}}

\DeclareMathOperator{\T}{\operatorname{T}}

\DeclareMathOperator{\fieldR}{\mathbb{R}}

\newcommand{\ve}[1]{\boldsymbol{#1}}

\newcommand{\exdi}[2]{\E_{#1} \left[#2\right]}

\newcommand{\expb}[1]{\exp{\left(#1\right)}}

\newcommand{\sign}[1]{\Sign{\left(#1\right)}}

\newcommand{\qfunc}[1]{\Q \left(#1\right)}

\DeclareMathAlphabet\mathbfcal{OMS}{cmsy}{b}{n}
\title{\huge Performance Analysis for Time-of-Arrival Estimation\\with Oversampled Low-Complexity 1-bit A/D Conversion}
\author{
Manuel S. Stein\\
\IEEEauthorblockA{Mathematics Department, Vrije Universiteit Brussel, Belgium}
E-mail: manuel.stein@vub.ac.be
\thanks{This research was supported by the German Academic Exchange Service (DAAD) with funds from the German Federal Ministry of Education and Research (BMBF) and the People Program (Marie Sk{\l}odowska-Curie Actions) of the European Union's Seventh Framework Program (FP7) under REA grant agreement 605728 (P.R.I.M.E. - Postdoctoral Researchers International Mobility Experience).}
}

\begin{document}
\maketitle
\begin{abstract}
Analog-to-digtial (A/D) conversion plays a crucial role when it comes to the design of energy-efficient and fast signal processing systems. As its complexity grows exponentially with the number of output bits, significant savings are possible when resorting to a minimum resolution of a single bit. However, then the nonlinear effect which is introduced by the A/D converter results in a pronounced performance loss, in particular for the case when the receiver is operated outside the low signal-to-noise ratio (SNR) regime. By trading the A/D resolution for a moderately faster sampling rate, we show that for time-of-arrival (TOA) estimation under any SNR level it is possible to obtain a low-complexity $1$-bit receive system which features a smaller performance degradation then the classical low SNR hard-limiting loss of $2/\pi$ ($-1.96$ dB). Key to this result is the employment of a lower bound for the Fisher information matrix which enables us to approximate the estimation performance for coarsely quantized receivers with correlated noise models in a pessimistic way.
\end{abstract}
\section{Introduction}
\label{sec:intro}
When it comes to the design of signal processing systems, it has been recently understood that A/D conversion forms a bottleneck at the receiver with respect to its power consumption and hardware complexity \cite{Walden99}. Therefore, in contrast to classical works on hard-limiting which where aiming at the minimization of the digital processing complexity \cite{Bennett48,Vleck66}, today the topic of $1$-bit quantization has found a vital revival due the necessity of reducing the analog sensing complexity \cite{Madsen00}-\cite{SteinBar_ICASSP2016}. This shift of attention to the analog sensor front-end is a consequence of Moore's law. While in the last four decades the computational capability per integrated circuit has approximately doubled every two years, the technological progress with respect to analog sensor hardware is much slower. Therefore, in the last years the design of receivers with low-complexity $1$-bit A/D conversion has been emphasized within the signal processing and communication community in order to meet the requirements of future wireless systems and standards which feature high signal bandwidth \cite{Heath16} and massive antenna arrays \cite{Jacobsson15}-\cite{SteinWSA16}.

Although $1$-bit A/D conversion at the receiver is usually associated with a performance loss of more than $-1.96$ dB \cite{Vleck66}, in this work we show that trading the resolution for a moderately higher sampling rate allows to design $1$-bit systems which outperform this classical benchmark for specific signal processing tasks. Obtaining this result requires to analyze the estimation accuracy with hard-limited Gaussian signal models featuring noise correlation. For such models the exact analytic representation of the likelihood function is an open mathematical problem \cite{Kedem80,Sinn11}. Here we circumvent this obstacle by a lower bound for the Fisher information measure \cite{Jarrett84} \cite{SteinTSP16}, resulting in a conservative approximation of the classical Cram\'er-Rao lower bound (CRLB) \cite{Rao45,Cram46}. Based on it, we visualize the asymptotic TOA estimation performance which can be achieved in different SNR scenarios with measurement data from hard-limiting receive sensors. Note that TOA estimation is a fundamental channel estimation problem with application in radar \cite{Weiss83}, radio-based positioning and synchronization \cite{Seco12}. 
\section{System Model}
\label{sec:model}
We assume a real-valued analog receive signal of the form
\begin{align}
\breve{y}(t)= \gamma \breve{x}(t-\tau)+\breve{\eta}(t),
\end{align}
with $\breve{x}(t)$ being a periodic pilot signal of structure
\begin{align}
\breve{x}(t) = \sum_{k=-\infty}^{+\infty} [\ve{c}]_{(1+\operatorname{mod}(k,M))} \breve{g}(t-kT_c),
\end{align}
where $\ve{c} \in \{-1,1\}^{M}$ is a binary sequence with $M$ elements and a chip frequency $f_c=\frac{1}{T_c}$. The duration of one pilot period is $T_o=MT_c$. For simplicity we assume that the transmit pulse is rectangular and band-limited to the bandwidth $B$,
\begin{align}
\breve{g}(t)=\frac{\operatorname{Si}\Big(2\pi B\big(t+\frac{T_c}{2}\big)\Big)-\operatorname{Si}\Big(2\pi B\big(t-\frac{T_c}{2}\big)\Big)}{\pi\sqrt{T_c}},
\end{align}
where we use the definition
\begin{align}
\operatorname{Si}(x)=\int_{0}^x\frac{\sin (u)}{u}{\rm d}u.
\end{align}
The parameter $\gamma\in\fieldR$ is associated with the attenuation and $\tau\in\fieldR$ with the time-delay of the propagation channel. The analog sensor signal $\breve{y}(t)$ is filtered by an ideal low-pass filter  
\begin{align}
H(\omega)=
\begin{cases} 
1 &\mbox{if } |\omega| \leq 2\pi B\\
0 & \mbox{else}
\end{cases}
\end{align}
with bandwidth $B$, such that the analog receive signal is
\begin{align}
{y}(t)&=\breve{y}(t) \ast h(t)\notag\\
&=  \gamma {x}(t-\tau)+{\eta}(t).
\end{align}
Assuming white Gaussian noise $\breve{\eta}(t)$ with constant power spectral density $\frac{N_{0}}{2}$, the temporal auto-correlation function of the additive noise after low-pass filtering
\begin{align}
r(t)=\int_{-\infty}^{\infty}{\eta}(\alpha){\eta}(\alpha-t){\rm d}\alpha,
\end{align}
can be characterized by the inverse Fourier transform
\begin{align}\label{autocorrelation:noise}
r(t)&=\frac{1}{2\pi}\int_{-\infty}^{\infty} \frac{N_{0}}{2} \left| H(\omega) \right|^2 e^{-{\rm j}\omega t}{\rm d}\omega\notag\\
&=B N_0 \operatorname{sinc}{(2Bt)},
\end{align}
where the sinc function is defined
\begin{align}
\operatorname{sinc}(x)=\frac{\sin{(\pi x)}}{\pi x}.
\end{align}
In the following we normalize the receive model such that 
\begin{align}
\frac{1}{T_o}\int_{0}^{T_o} |{x}(t)|^2 {\rm d}t=1
\end{align}
and $r(0)=1$. Therefore, the attenuation parameter 
\begin{align}
\gamma=\frac{1}{\sqrt{B N_0}}=\sqrt{\text{SNR}}
\end{align}
stands in relation to the receive SNR. 
The analog sensor signal $y(t)$ is discretized in time at a sampling frequency of $f_s=\frac{1}{T_s}$, such that the digital receive signal
\begin{align}
\ve{y}=\gamma \ve{x}(\tau)+\ve{\eta},
\end{align}
with $\ve{y}, \ve{x}(\tau),\ve{\eta}\in\fieldR^{N}$ and vector entries
\begin{align}
[\ve{y}]_i&=y\big((i-1)T_s\big),\\
[\ve{x}(\tau)]_i&=x\big((i-1)T_s-\tau\big),\\
[\ve{\eta}]_i&=\eta\big((i-1)T_s\big),
\end{align}
is obtained. Due to the form of the noise auto-correlation function \eqref{autocorrelation:noise}, the entries of the normalized covariance matrix 
\begin{align}
\ve{R}_{\ve{\eta}}=\exdi{\ve{\eta}}{\ve{\eta}\ve{\eta}^{\T}}
\end{align}
are given by
\begin{align}
[\ve{R}_{\ve{\eta}}]_{ij}=\operatorname{sinc}{(2BT_s\left| i-j \right|)}.
\end{align}
It is observed that white noise, i.e., $\ve{R}_{\ve{\eta}}= \ve{I}_N$, is only obtained if the relation $f_{s}={2B}$ between the sampling rate and the receive filter bandwidth is satisfied exactly. 
For convenience, in the following we write the receive signal model
\begin{align}\label{system:model:ideal}
\ve{y}&= \ve{s}(\ve{\theta})+\ve{\eta}
\end{align}
and summarize the channel parameters by 
\begin{align}
\ve{\theta}=\begin{bmatrix} \gamma &\tau \end{bmatrix}^{\T}.
\end{align}
In order to model a receiver with low-complexity $1$-bit A/D conversion we use
\begin{align}\label{system:model:sign}
\ve{z}= \sign{\ve{y}},
\end{align}
where $\sign{\cdot}$ is the element-wise signum function. Note that \eqref{system:model:sign} models an A/D conversion without feedback loop. This separates low-complexity $1$-bit A/D conversion from the sigma-delta modulation approach, where a single comparator with feedback is operated in a highly oversampled mode \cite{Aziz96}. 
\section{Channel Estimation Performance}
\label{sec:performance}
The signal processing task considered here is to calculate the maximum-likelihood (ML) estimate of the parameters
\begin{align}\label{definition:mle:hardlim}
\ve{\hat{\theta}}(\ve{z})=\arg \max_{\ve{\theta}\in\ve{\Theta}} \ln p(\ve{z};\ve{\theta})
\end{align}
from the hard-limited receive signal \eqref{system:model:sign}. When analyzing the achievable accuracy with the procedure \eqref{definition:mle:hardlim}, we use the ideal system \eqref{system:model:ideal} as a benchmark, for which the ML estimator is calculated from the unquantized receive signal
\begin{align}\label{definition:mle:ideal}
\ve{\hat{\theta}}(\ve{y})=\arg \max_{\ve{\theta}\in\ve{\Theta}} \ln p(\ve{y};\ve{\theta}).
\end{align}
For unbiased processing algorithms the performance is in general lower bounded by the CRLB \cite{Rao45,Cram46}
\begin{align}
\exdi{\ve{z};\ve{\theta}}{\big(\ve{\hat{\theta}}(\ve{z})-\ve{\theta}\big) \big(\ve{\hat{\theta}}(\ve{z})-\ve{\theta}\big)^{\T}} &\succeq \ve{F}^{-1}_{\ve{z}} (\ve{\theta}),\label{crlb:1bit}\\
\exdi{\ve{y};\ve{\theta}}{\big(\ve{\hat{\theta}}(\ve{y})-\ve{\theta}\big) \big(\ve{\hat{\theta}}(\ve{y})-\ve{\theta}\big)^{\T}} &\succeq \ve{F}^{-1}_{\ve{y}} (\ve{\theta}),\label{crlb:ideal}
\end{align}
where the Fisher information matrices are defined \cite{Trees07}
\begin{align}
\ve{F}_{\ve{z}} (\ve{\theta})&=\exdi{\ve{z};\ve{\theta}}{ \bigg(\frac{\partial \ln p(\ve{z};\ve{\theta}) }{\partial \ve{\theta}} \bigg)^{\T} \frac{\partial \ln p(\ve{z};\ve{\theta}) }{\partial \ve{\theta}}},\label{fim:1bit}\\
\ve{F}_{\ve{y}} (\ve{\theta})&=\exdi{\ve{y};\ve{\theta}}{ \bigg(\frac{\partial \ln p(\ve{y};\ve{\theta}) }{\partial \ve{\theta}} \bigg)^{\T} \frac{\partial \ln p(\ve{y};\ve{\theta}) }{\partial \ve{\theta}}}\label{fim:ideal}.
\end{align}
Note that asymptotically the ML estimators \eqref{definition:mle:hardlim} and \eqref{definition:mle:ideal} are unbiased and obtain equality in \eqref{crlb:1bit} and \eqref{crlb:ideal} \cite{Kay93}. While for the ideal receiver, with
\begin{align}
\frac{\partial \ve{s}(\ve{\theta}) }{\partial\ve{\theta}}&= 
\begin{bmatrix}\frac{\partial \ve{s}(\ve{\theta}) }{\partial\gamma} & \frac{\partial \ve{s}(\ve{\theta}) }{\partial\tau} \end{bmatrix}\notag\\
&=\begin{bmatrix}\ve{x}(\tau) & \gamma\frac{\partial \ve{x}(\tau)}{\partial \tau} \end{bmatrix}
\end{align}
and
\begin{align}
\left[\frac{\partial \ve{x}(\tau)}{\partial \tau} \right]_{i} &=-\frac{\mathrm d x(t)}{\mathrm d t}\Big|_{t=(i-1)T_s-\tau},
\end{align}
the Fisher information measure is obtained simply by
\begin{align}
\ve{F}_{\ve{y}} (\ve{\theta})= \bigg( \frac{\partial \ve{s}(\ve{\theta})}{\partial \ve{\theta}} \bigg)^{\T} \ve{R}^{-1}_{\ve{\eta}}  \frac{\partial \ve{s}(\ve{\theta})}{\partial \ve{\theta}},
\end{align}
for the model with hard-limiting \eqref{system:model:sign}, the likelihood function $p(\ve{z};\ve{\theta})$ required in \eqref{fim:1bit} is non-trivial for cases where $N>4$. This is due to the fact that the characterization of the orthant probability (multivariate version of the Q-function) is in general an open mathematical problem \cite{Sinn11}. Therefore, we employ a lower bound for the Fisher information matrix \cite{Jarrett84} \cite{SteinTSP16}
\begin{align}\label{lower:bound:fisher}
\ve{F}_{\ve{z}} (\ve{\theta}) \succeq \ve{\tilde{F}}_{\ve{z}} (\ve{\theta}) 
\end{align}
with
\begin{align}\label{pessimistic:fisher:matrix}
\ve{\tilde{F}}_{\ve{z}} (\ve{\theta}) = \bigg(\frac{\partial \ve{\mu}_{\ve{\phi}}(\ve{\theta}) }{\partial\ve{\theta}} \bigg)^{\rm T} \ve{R}_{\ve{\phi}}^{-1}(\ve{\theta}) \bigg(\frac{\partial \ve{\mu}_{\ve{\phi}}(\ve{\theta}) }{\partial\ve{\theta}} \bigg),
\end{align}
where the required mean and covariance are
\begin{align}
\ve{\mu}_{\ve{\phi}}(\ve{\theta})&=\exdi{\ve{z};\ve{\theta}}{\ve{\phi}(\ve{z})},\label{mean:aux:stat}\\
\ve{R}_{\ve{\phi}}(\ve{\theta})&=\exdi{\ve{z};\ve{\theta}}{\ve{\phi}(\ve{z})\ve{\phi}^{\rm T}(\ve{z})}-\ve{\mu}_{\ve{\phi}}(\ve{\theta})\ve{\mu}_{\ve{\phi}}^{\rm T}(\ve{\theta}),\label{covariance:aux:stat}
\end{align}
while $\ve{\phi}(\ve{z}): \fieldR^N \to \fieldR^{L}$ is an arbitrary transformation. Note that the information bound \eqref{lower:bound:fisher} can be derived after replacing the likelihood $p(\ve{z};\ve{\theta})$ by an equivalent model $\tilde{p}(\ve{z};\ve{\theta})$ within the exponential family \cite{SteinPhD}. Here we use identity $\ve{\phi}(\ve{z})=\ve{z}$, such that \eqref{mean:aux:stat} can be calculated element-wise by \cite{SteinWCL15}
\begin{align}\label{first:moment:hardlim}
[\ve{\mu}_{\ve{z}}(\ve{\theta})]_i&=p\big([\ve{z}]_i =1; \ve{\theta}\big)-p\big([\ve{z}]_i = -1; \ve{\theta}\big)\notag\\
&=1-2\qfunc{ \frac{  [\ve{s}(\ve{\theta})]_{i}}{\sqrt{[\ve{R}_{\ve{\eta}}]_{ii}}}}
\end{align}
with $\qfunc{\cdot}$ denoting the Q-function
\begin{align}
\qfunc{x}=\frac{1}{\sqrt{2\pi}} \int_{x}^{\infty} \exp{\left(-\frac{u^2}{2}\right)} {\rm d} u.
\end{align}
For the covariance matrix \eqref{covariance:aux:stat}, the diagonal elements are
\begin{align}
[\ve{R}_{\ve{z}}(\ve{\theta})]_{ii}&=1-[\ve{\mu}_{\ve{z}}(\ve{\theta})]_i^2,
\end{align}
while the off-diagonal entries are calculated
\begin{align}
[\ve{R}_{\ve{z}}(\ve{\theta})]_{ij}&=4\Psi_{ij}(\ve{\theta}) -\big(1-[\ve{\mu}_{\ve{z}}(\ve{\theta})]_i\big) \big(1-[\ve{\mu}_{\ve{z}}(\ve{\theta})]_j\big),
\end{align}
where $\Psi_{ij}(\ve{\theta}) $ is the cumulative density function (CDF) of the bivariate Gaussian distribution
\begin{align}
\mathcal{N}\Bigg(\begin{bmatrix} 0\\ 0 \end{bmatrix},\begin{bmatrix} [\ve{R}_{\ve{\eta}}]_{ii} &[\ve{R}_{\ve{\eta}}]_{ij}\\ [\ve{R}_{\ve{\eta}}]_{ji} &[\ve{R}_{\ve{\eta}}]_{jj} \end{bmatrix} \Bigg),
\end{align}
with upper integration border $\begin{bmatrix} - [\ve{s}(\ve{\theta})]_{i} &- [\ve{s}(\ve{\theta})]_{j}\end{bmatrix}^{\rm T}$. With
\begin{align}
\frac{\partial \qfunc{x}}{\partial x}= -\frac{1}{\sqrt{2\pi}} \exp{\left(-\frac{x^2}{2}\right)},
\end{align}
the derivative of \eqref{first:moment:hardlim} is found element-wise
\begin{align}
\Bigg[ \frac{\partial \ve{\mu}_{\ve{z}}(\ve{\theta}) }{\partial\ve{\theta}} \Bigg]_{ij}= \frac{2  \expb{- \frac{ {s}_{i}^2(\ve{\theta})}{{2 [\ve{R}_{\ve{\eta}}]_{ii}}} } }{ \sqrt{2 \pi [\ve{R}_{\ve{\eta}}]_{ii} } } \Bigg[ \frac{\partial \ve{s}(\ve{\theta}) }{\partial\ve{\theta}} \Bigg]_{ij}.
\end{align}
The performance gap between the ideal receiver \eqref{definition:mle:ideal} and the $1$-bit system \eqref{definition:mle:hardlim} with respect to the estimation of both channel parameters $\gamma$ and $\tau$ can be characterized by the ratios
\begin{align}
\chi_{\gamma}(\ve{\theta}) &=  \frac{\big[\ve{F}_{\ve{y}}^{-1}(\ve{\theta})\big]_{11}}{\big[\ve{\tilde{F}}_{\ve{z}}^{-1}(\ve{\theta})\big]_{11}},\label{qloss:attenuation}\\
\chi_{\tau} (\ve{\theta})&=  \frac{\big[\ve{F}_{\ve{y}}^{-1}(\ve{\theta})\big]_{22}}{\big[\ve{\tilde{F}}_{\ve{z}}^{-1}(\ve{\theta})\big]_{22}}.\label{qloss:delay}
\end{align}
\section{Results}
\label{sec:results}
For visualization of the results, we consider a GPS-like setup \cite{GPS_spec}, with $M=1023$ random binary pilot symbols and a chip frequency $f_c=\frac{1}{T_c}=1.023$ MHz, such that the symbol duration is $T_c=977.52$ ns and $T_o=1$ ms. The sampling rate is set to $f_{s}={2B}\kappa$ with the oversampling factor $\kappa\geq1$ while the one-sided bandwidth of the analog pre-filter is fixed to $B=1.023$ MHz. For the case $\kappa=1$ this setup results in $N=2046$ digital receive samples. 
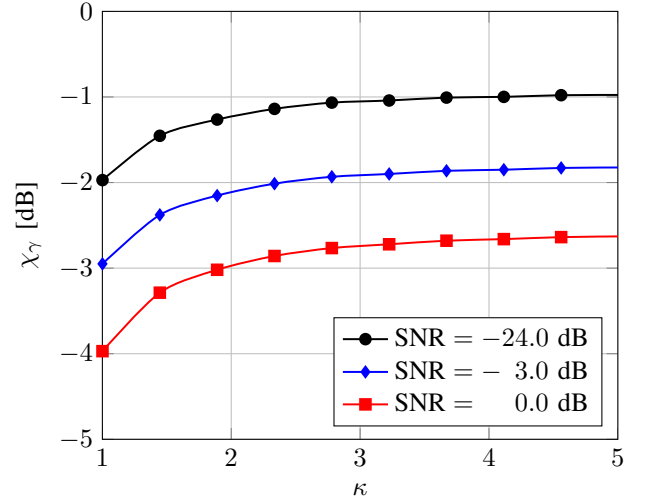
\begin{figure}[h!]
\centering
\begin{tikzpicture}[scale=1]

  	\begin{axis}[ylabel=$ \chi_\gamma\text{ [dB]}$,
  			xlabel=$\kappa$,
			grid,
			ymin=-5.0,
			ymax=0,
			xmin=1,
			xmax=5,
			legend pos=south east]

			\addplot[black, style=solid, line width=0.75pt,smooth,every mark/.append style={solid}, mark=otimes*, mark repeat=1] table[x index=0, y index=1]{QLoss_SamplingRate_m24dB.txt};
			\addlegendentry{$\text{SNR}=-24.0$ dB}
			
			\addplot[blue, style=solid, line width=0.75pt,smooth,every mark/.append style={solid}, mark=diamond*, mark repeat=1] table[x index=0, y index=1]{QLoss_SamplingRate_m3dB.txt};
			\addlegendentry{$\text{SNR}=-\phantom{1}3.0$ dB}
			
			\addplot[red, style=solid, line width=0.75pt,smooth,every mark/.append style={solid}, mark=square*, mark repeat=1] table[x index=0, y index=1]{QLoss_SamplingRate_m0dB.txt};
			\addlegendentry{$\text{SNR}=\phantom{-1}0.0$ dB}

	\end{axis}
	
\end{tikzpicture}
\caption{Performance $\hat{\gamma}(\ve{z})$ vs. Oversampling $\kappa$}
\label{fig:oversampling:gamma}
\end{figure}
Fig. \ref{fig:oversampling:gamma} shows the $1$-bit quantization loss \eqref{qloss:attenuation} for the attenuation parameter $\gamma$ as a function of the oversampling factor $\kappa$ in three different SNR scenarios. Without oversampling, i.e., $\kappa=1$, the classical result of $-1.96$ dB is obtained for the low SNR regime ($\text{SNR}=-24.0$ dB), while the loss is more pronounced at higher SNR values. Oversampling the receive signal allows to recover approximately $1$ dB of the initial quantization loss in all considered SNR scenarios. For example, for the setup with $\text{SNR}=-24.0$ dB the loss in accuracy reduces to $-0.98$ dB by oversampling with $\kappa=5$. A similar effect is observed for the medium SNR setting ($\text{SNR}=0.0$ dB) where oversampling allows to diminish the performance gap from $-3.97$ dB to $-2.62$ dB by sampling at a higher rate. Note that the ideal receive system \eqref{definition:mle:ideal} does not benefit from oversampling as, due to the sampling theorem \cite{Luke99}, the analog receive signal $y(t)$ can be reconstructed without error from the samples $\ve{y}$ for all configurations with $\kappa\geq1$.
\begin{figure}[h!]
\centering
\begin{tikzpicture}[scale=1]

  	\begin{axis}[ylabel=$\chi_\tau\text{ [dB]}$,
  			xlabel=$\kappa$,
			grid,
			ymin=-3.0,
			ymax=-0.5,
			xmin=1,
			xmax=5,
			legend pos=south east]

			\addplot[black, style=solid, line width=0.75pt,smooth,every mark/.append style={solid}, mark=otimes*, mark repeat=1] table[x index=0, y index=2]{QLoss_SamplingRate_m24dB.txt};
			\addlegendentry{$\text{SNR}=-24.0$ dB}
			
			\addplot[blue, style=solid, line width=0.75pt,smooth,every mark/.append style={solid}, mark=diamond*, mark repeat=1] table[x index=0, y index=2]{QLoss_SamplingRate_m6dB.txt};
			\addlegendentry{$\text{SNR}=-\phantom{1}6.0$ dB}
			
			\addplot[red, style=solid, line width=0.75pt,smooth,every mark/.append style={solid}, mark=square*, mark repeat=1] table[x index=0, y index=2]{QLoss_SamplingRate_m0dB.txt};
			\addlegendentry{$\text{SNR}=\phantom{-1}0.0$ dB}
		
	\end{axis}
	
\end{tikzpicture}
\caption{Performance $\hat{\tau}(\ve{z})$ vs. Oversampling $\kappa$}
\label{fig:oversampling:tau}
\end{figure}
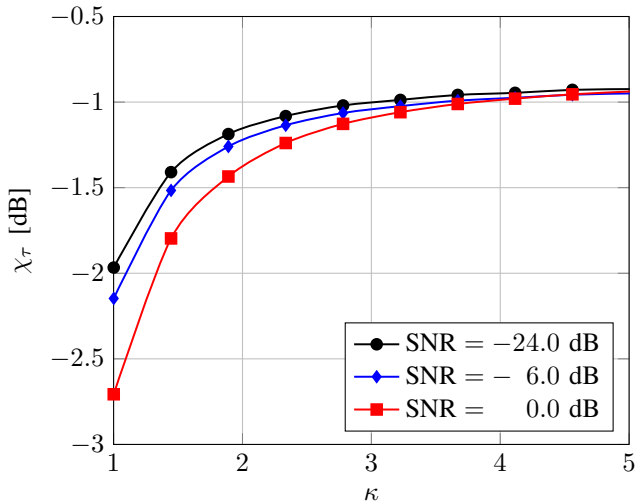

An interesting result is obtained when analyzing the $1$-bit quantization loss \eqref{qloss:delay} for the time-delay parameter $\tau$ as a function of the oversampling factor $\kappa$ (see Fig. \ref{fig:oversampling:tau}). While for the low SNR regime the time-delay accuracy loss \eqref{qloss:delay} shows a behavior similar to the performance gap of the attenuation parameter $\gamma$, in the medium SNR regime we observe a significant performance improvement when oversampling the receive signal. In the medium SNR situation where $\text{SNR}=0.0$ dB, the initial $1$-bit quantization loss without oversampling is $-2.70$ dB while with oversampling with $\kappa=5$ a gap of only $-0.99$ dB is reached. Taking into account that the power dissipation $P_{\text{ADC}}(b, f_s)$ of an ADC scales
\begin{align}
P_{\text{ADC}}(b, f_s)\approx\beta_{\text{ADC}} (2^b-1)f_s,
\end{align}
where $\beta_{\text{ADC}}$ is a constant dependent on the particular ADC technology and $b$ the output resolution, it can be concluded that the $1$-bit receiver can be operated at $\kappa=3$ with a hardware complexity similar to a $2$-bit ADC running at $\kappa=1$. Note that this is a conservative statement as in comparison to a $2$-bit converter the low-complexity $1$-bit ADC does not require an automatic gain control (AGC). From Fig. \ref{fig:oversampling:tau} we can therefore see that the $1$-bit TOA loss can be made smaller than $-1.10$ dB independently of the SNR, when normalizing to the same A/D complexity. This is significantly less than the classical benchmark of $-1.96$ dB and shows that oversampling is a simple but effective approach in order to compensate the loss introduced by a low-complexity $1$-bit ADC. 
\section{Conclusion}
\label{sec:conclusion}
We have analyzed the channel estimation performance when A/D conversion with an output resolution of a single bit is performed at the receiver. With a pessimistic approximation of the Fisher information measure, an asymptotic performance analysis based on the classical CRLB was presented which includes the case where oversampling is used and the signal model therefore exhibits correlated noise. The obtained results show that in particular the accuracy of the TOA channel parameter can be significantly increased through oversampling. This confirms that low-complexity $1$-bit A/D conversion at the receiver is an interesting system design option for future wireless systems, in particular for applications like radar, radio-based positioning and synchronization which require a high-resolution estimate of the TOA channel parameter at small hardware cost. 

\end{document}